\documentstyle[aps,multicol,epsf]{revtex}

\begin{document}
\draft

\title{Packing of Compressible Granular Materials} 

\author{Hern\'an A. Makse, David L.
Johnson, and Lawrence M. Schwartz}

\address{ Schlumberger-Doll Research, Old Quarry Road, Ridgefield, CT
06877} 
\date{\today} 
\maketitle
\begin{abstract}

3D Computer simulations and experiments are employed to study random
packings of compressible spherical grains under external confining
stress.  Of particular interest is the rigid ball limit, which we
describe as a continuous transition in which the applied stress
vanishes as $(\phi-\phi_c)^\beta$, where $\phi$ is the (solid phase)
volume density.  This transition coincides with the onset of shear
rigidity.  The value of $\phi_c$ depends, for example, on whether the
grains interact via only normal forces (giving rise to random close
packings) or by a combination of normal and friction generated
transverse forces (producing random loose packings).  In both cases,
near the transition, the system's response is controlled by localized
force chains.  As the stress increases, we characterize the system's
evolution in terms of (1) the participation number, (2) the average
force distribution, and (3) visualization techniques.
\end{abstract}


\begin{multicols}{2}

Dense packings of spherical particles are an important starting point
for the study of physical systems as diverse as simple liquids,
metallic glasses, colloidal suspensions, biological systems, and
granular matter \cite{bernal,finney,berryman,ibm,torquato}.  In the
case of liquids and glasses, finite temperature molecular dynamics
(MD) studies of hard sphere models have been particularly important.
Here one finds a first order liquid-solid phase transition as the
solid phase volume fraction, $\phi$, increases. Above the freezing
point, a metastable disordered state can persist until
$\phi\to\phi_{\mbox{\scriptsize RCP}}^-$ \cite{torquato}, where
$\phi_{\mbox{\scriptsize RCP}}$ is the density of random close packing
(RCP)--- the densest possible random packing of hard spheres.

This Letter is concerned with the non-linear elastic properties of
granular packings.  Unlike glasses and amorphous solids, this is a
zero temperature system in which the interparticle forces are both
non-linear, and path (i.e., history) dependent.  [Because these forces
are purely repulsive, mechanical stability is achieved only by imposing
external stress.]  The structure of packing depends in detail on the
forces acting between the grains during rearrangement of grains;
indeed, different rearrangement protocols can lead to either RCP or
random loose packed (RLP) systems.

In the conventional continuum approach to this problem, the granular
material is treated as an elasto-plastic medium \cite{nedderman}.  
However, this
approach has been challenged by recent authors \cite{bouchaud}
who argue that granular
packings represents a new kind of {\it fragile} matter and that more
exotic methods, e.g., the fixed principal axis ansatz, are required to
describe their internal stress distributions.  These new continuum
methods are complemented by microscopic studies based on either 
contact dynamics simulations of {\it rigid} spheres or statistical
models, such as the q-model, which makes no attempt to take account of
the character of the inter-grain forces \cite{chicago,radjai}.

In our view, a proper description of the stress state in granular
systems must take account of the fact that the individual grains are
deformable.  We report here on a 3D study of deformable spheres
interacting via Hertz-Mindlin contact forces.  Our simulations cover
four decades in the applied pressure and allow us to understand the
regimes in which the different theoretical approaches described above
are valid.  Since the grains in our simulations are deformable, the
volume fraction can be increased above the hard sphere limit and we
are able to study the approach to the RCP and RLP
states from this realistic perspective.
 Within this framework, the rigid
grain limit is described as a continuous phase transition where the
order parameter is the applied stress, $\sigma$, which vanishes
continuously as $(\phi-\phi_c)^\beta$.  Here $\phi_c$ is the critical
volume density, and $\beta$ is the corresponding critical exponent.
We emphasize that the fragile state corresponding
to rigid grains is reached by looking at the limit $\phi\to\phi^+_c$ from 
above.


Of particular importance is the fact that $\phi_c$ depends on the type
of interaction between the grains.  If the grains interact via normal
forces only \cite{bubbles}, they slide and rotate freely mimicking the
rearrangements of grains during shaking in experiments
\cite{bernal,finney,berryman}.  We then obtain the RCP value $\phi_c=
0.634(4)(\approx\phi_{\mbox{\scriptsize RCP}})$.  By contrast, if the
grains interact by combined normal and friction generated transverse
forces, we get RLP \cite{ibm} at the critical point with
$\phi_c=0.6284(2)<\phi_{\mbox{\scriptsize RCP}}$.  The power-law
exponents characterizing the approach to $\phi_c$ are not universal
and depend on the strength of friction generated shear forces.

Our results indicate that the transitions at both RCP or RLP are
driven by localized force chains. Near the critical
density there is a percolative fragile structure which we characterize
by the participation number (which measures localization of force
chains), the probability distribution of forces, and also by
visualization techniques.  
A subset of our results are experimentally
verified using carbon paper measurements to study force
distributions in the granular assembly.   We also consider in some
detail the relationship between our work and recent experiments in
2D Couette geometries \cite{behringer}.


\begin{figure}
\centerline{
\hbox{ 
\epsfxsize=4.cm
\epsfbox{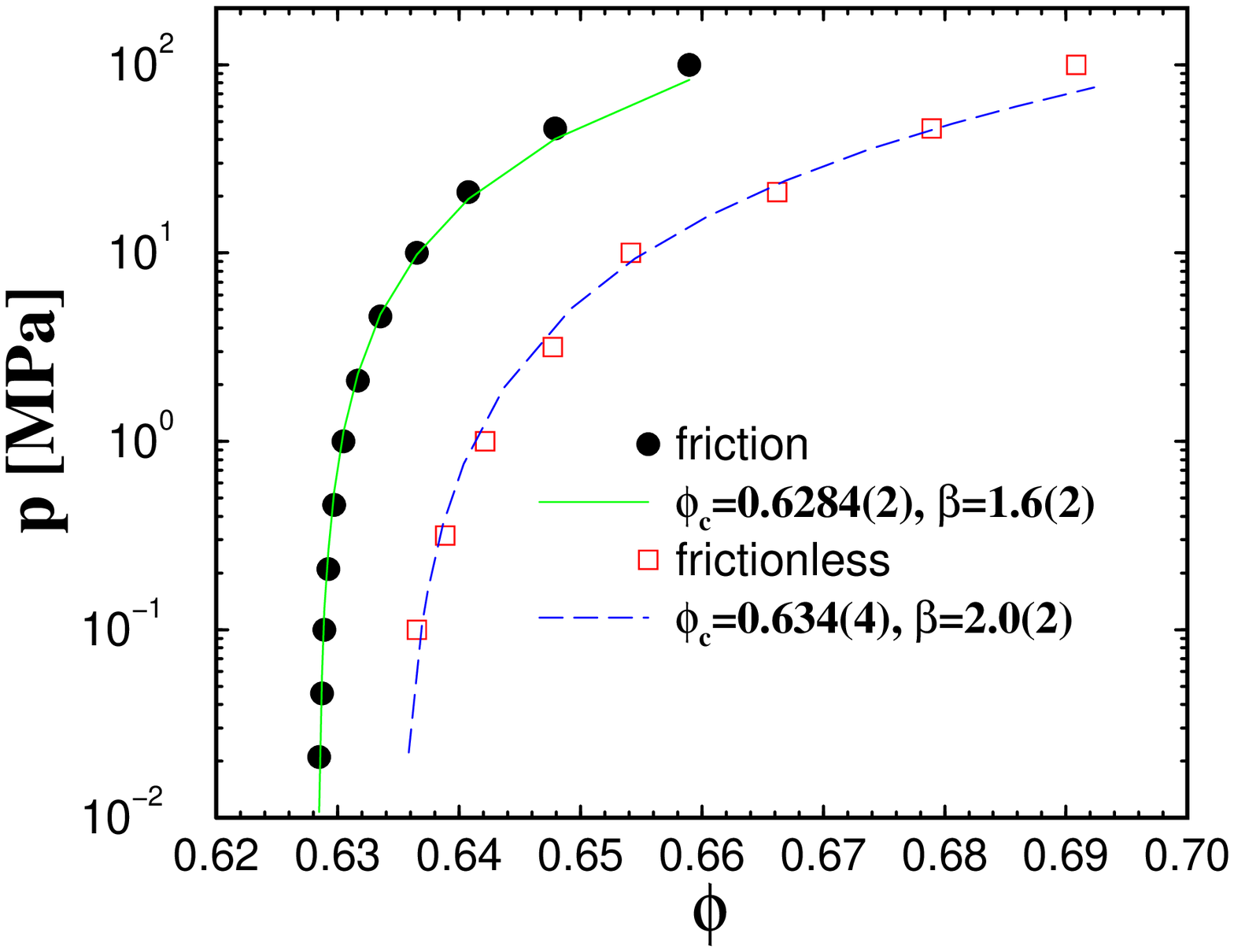}
\epsfxsize=4.cm
\epsfbox{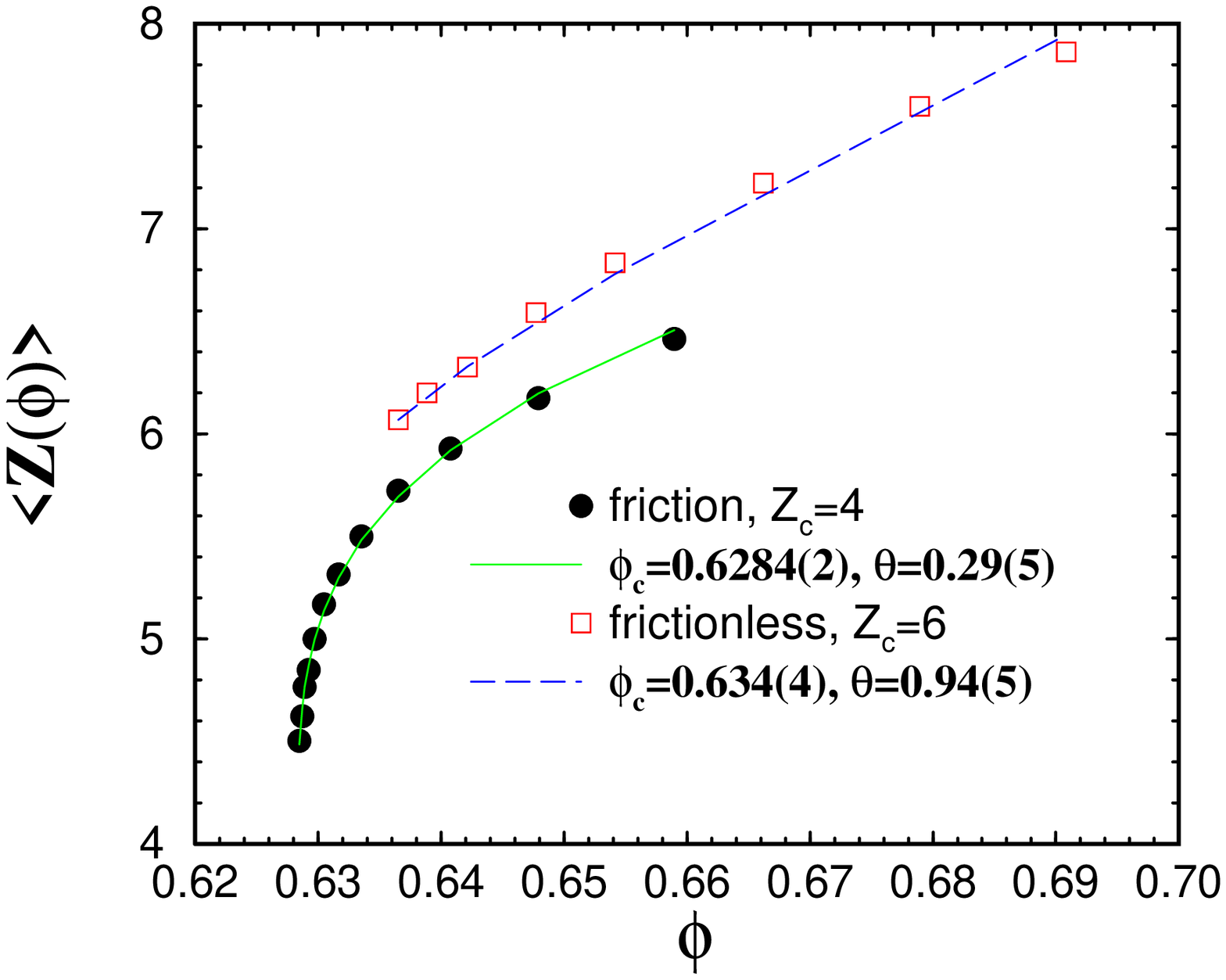}
}}
\narrowtext
\caption{(a) Confining stress and (b)
average coordination number
as a function of volume fraction for friction and frictionless balls.}
\label{pressure-Z-pi}
\end{figure}

{\it Numerical Simulations}: To better understand the behavior of real
granular materials, we perform granular dynamics simulations of
unconsolidated packings of deformable spherical glass beads using the
Discrete Element Method developed by Cundall and Strack
\cite{cundall}.  Unlike previous work on rigid grains, we work with a
system of deformable elastic grains interacting via normal and
tangential Hertz-Mindlin forces plus viscous dissipative forces
\cite{johnson}.  The grains have shear modulus 29 GPa,
Poisson's ratio 0.2 and radius 0.1 mm.

Our simulations employ periodic boundary conditions and begin with a
gas of 10000 non-overlapping spheres located at random positions in a
cube 4 mm on a side.  Generating a mechanically stable packing is not
a trivial task \cite{torquato}.  At the outset, a series of
strain-controlled isotropic compressions and expansions are applied
until a volume fraction slightly below the critical density.  At this
point the system is at zero pressure and zero coordination number.  We
then compress along the $z$ direction,
until the system equilibrates at a desired vertical stress $\sigma$
 and a non-zero average coordination number $\langle Z \rangle$.

Figure \ref{pressure-Z-pi}a shows the behavior of the stress as a
function of the volume fraction. We find that the pressure vanishes at
a critical $\phi_c=0.6284(2)$. Although we cannot rule out a
discontinuity in the pressure at $\phi_c$--- as we could expect for a
system of hard spheres--- our results indicates that the transition is
continuous and the behavior of the pressure can be fitted to a power
law form
\begin{equation}
\sigma \sim (\phi - \phi_c)^{\beta},
\label{p}
\end{equation}
where $\beta = 1.6(2)$.  Our 3D results contrast with recent
experiments of slowly sheared grains in 2D Couette geometries
\cite{behringer} where a faster than exponential approach to $\phi_c$
was found, while they agree qualitative with similar continuous
transition found in compressed emulsions and foams \cite{bubbles}.

Figure \ref{pressure-Z-pi}b shows the behavior of the mean
coordination number, $ \langle Z \rangle$, as a function of $\phi$.
We find
\begin{equation}
 \langle Z \rangle - Z_c \sim (\phi - \phi_c)^{\theta},
\label{z}
\end{equation}
where $Z_c=4$ is a minimal coordination number, and $\theta=0.29(5)$
is a critical exponent.  At criticality the system is very loose and
fragile with a very low coordination number. The value of $Z_c$ can be
understood in term of constraint arguments as discussed in
\cite{edwards}; in the rigid ball limit, for a disordered system with
both normal and transverse forces, we find $Z_c = D+1 = 4$
\cite{edwards}.  As we compress the system more contacts are created,
providing more constraints so that the forces become overdetermined.

We notice that $\phi_c$ obtained for this system is considerably lower
than the best estimated value at RCP \cite{berryman}, 
$\phi_{\mbox{\scriptsize
RCP}}=0.6366(4)$ obtained by Finney 
\cite{finney} using ball bearings.  This latter value is obtained by
carefully vibrating the system and letting the grains settle into the
most compact packing. Numerically, this is achieved by allowing the
grains reach the state of mechanical equilibrium interacting only via
normal forces. By removing the transverse forces, grains can slide
freely and find most compact packings than with transverse forces.
Numerically we confirm this by equilibrating the system at zero
transverse force.  The critical packing fraction found in
this way is $\phi_c=0.634(4)$($ \approx \phi_{\mbox{\scriptsize RCP}}$
within error bars).  The stress behaves as in Eq. (\ref{p}) but with a
different exponent $\beta=2.0(2)$ (Fig. \ref{pressure-Z-pi}a).  At the
critical volume fraction the average coordination number is now $
Z_c=6$ [and $\theta=0.94(5)$, Fig. \ref{pressure-Z-pi}b], which again
can be understood using constraint arguments which would give a
minimal coordination number equal to 2D for frictionless rigid balls
\cite{edwards}.

\begin{figure}
\centerline{ 
\hbox{
\epsfxsize=4.cm \epsfbox{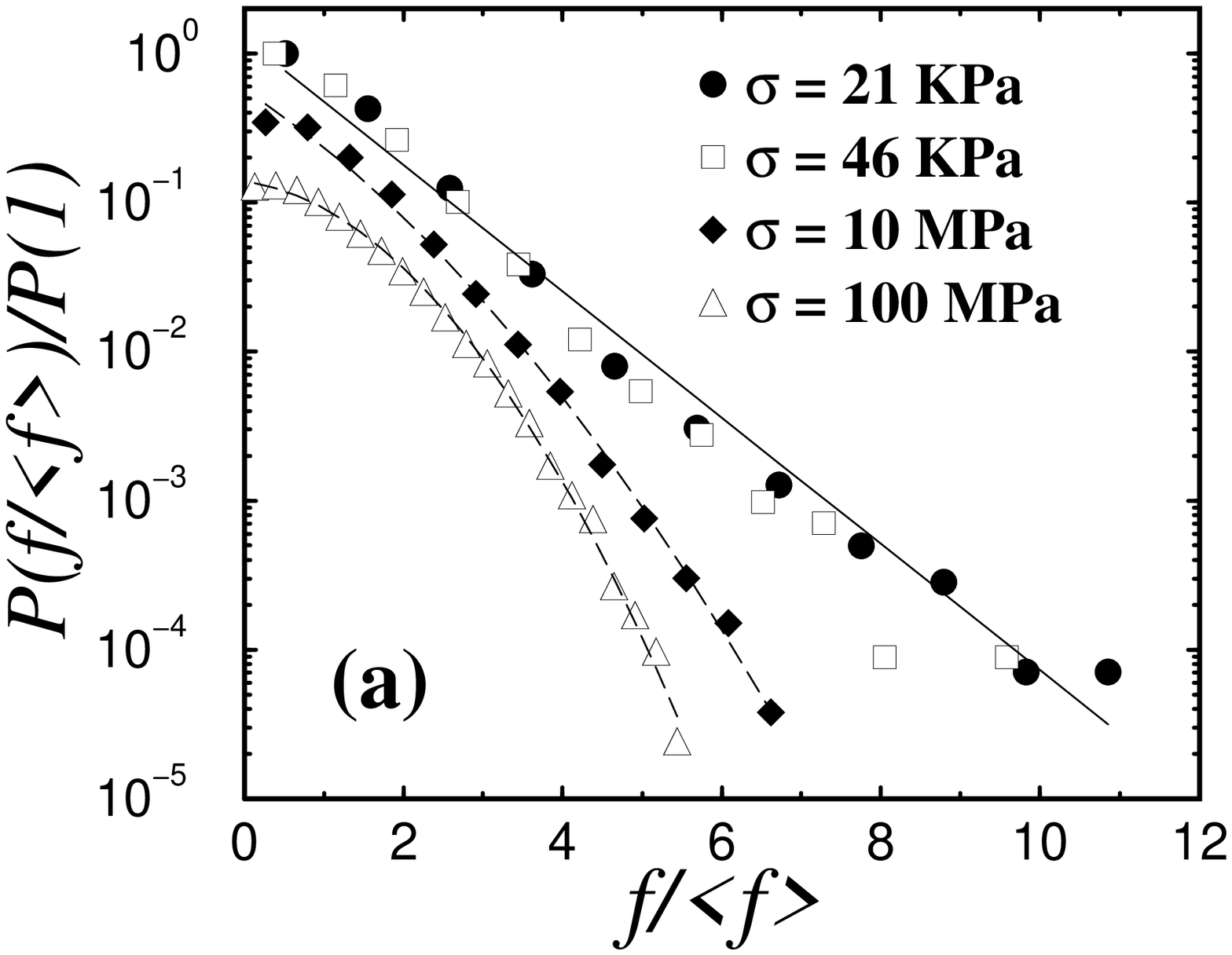} 
\epsfxsize=4.cm \epsfbox{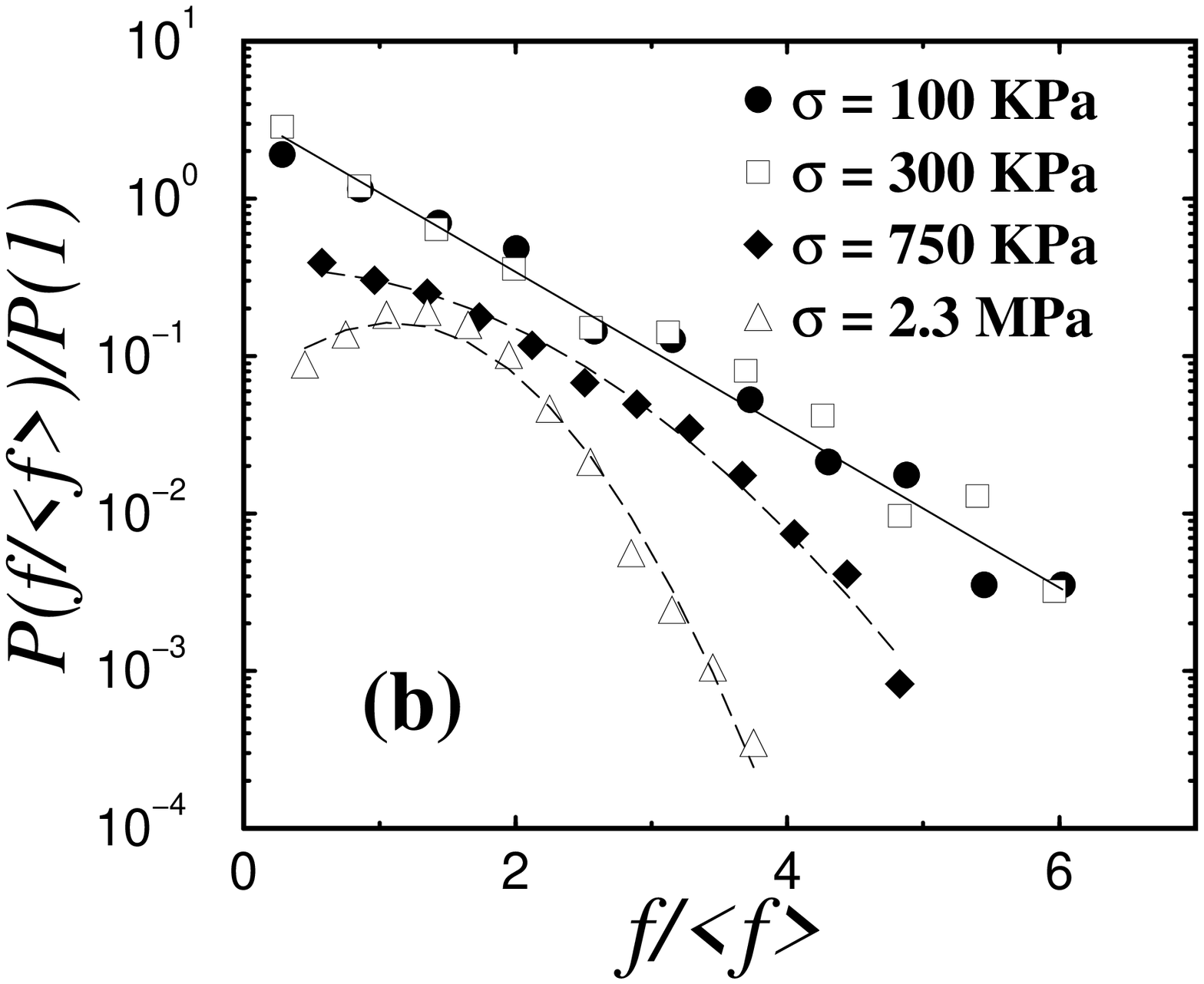} } }
\centerline{  
\epsfxsize=5.cm  \epsfbox{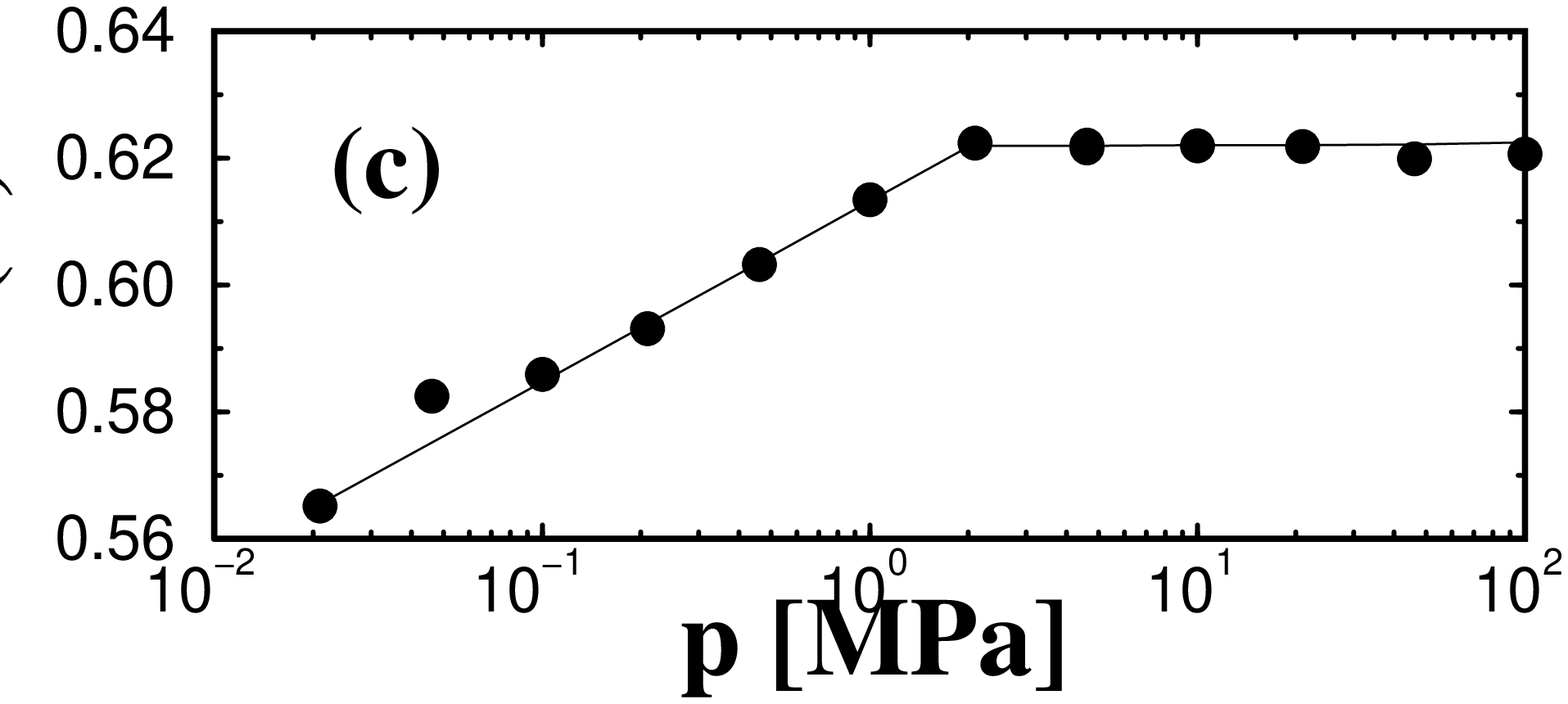} }
\narrowtext
\caption{Distribution of forces for different confining  stresses $\sigma$
obtained (a) in
the numerical simulations of friction balls, and (b) in the carbon paper
experiments.  The straight solid lines are fittings to
exponential forms and the dashed lines are fittings to Gaussian forms.
In both graphs we shift down the distributions corresponding to the
two larger stresses for clarity.
(c) Participation number versus external stress for the same system as in 
(a).}
\label{distri}
\end{figure}
\vspace{-.5cm}
We conclude that the value $\phi_c \approx
0.6288<\phi_{\mbox{\scriptsize RCP}}$ found with transverse forces
corresponds to the RLP limit, experimentally achieved by pouring balls
in a container but without allowing for further rearrangements \cite{ibm}.
Experimentally, stable loose 
packings with $\phi$ as low as 0.60 have been
found \cite{ibm}.  In our simulations,  $\phi_c$ lower than $0.6288$ can
be obtained by increasing the strength of the tangential forces.  This
is in agreement with experiments of Scott and Kilgour \cite{scott2}
who found that the maximum packing density of spheres decreases with
the surface roughness (friction) of the balls.

While previous studies characterized RCP's and RLP's by using radial
distribution functions and Voronoi constructions \cite{finney}, we
take a different approach which allow us to compare our results
directly with recent work on force transmissions in granular matter.
Previous studies of granular media indicate that, for forces greater
than the average value, the distribution of inter-grain contact forces
is exponential \cite{chicago,radjai}.
In addition, photo-elastic
visualization experiments and simulations \cite{force_chains,chicago}
show that contact forces are strongly localized along ``force chains''
which carry most of the applied stress.
The existence of force chains and exponential force distributions are
thought to be intimately related.

Here we analyze this scenario in the entire range of pressures: from
the $\phi_c$ limit and up.  Figure \ref{distri}a shows the force
distribution obtained in the simulations with friction balls.  At low
stress, the distribution is exponential in agreement with previous
experiments and models.
When the system is compressed further, we find a gradual transition to
a Gaussian force distribution. We observe a similar transition in our
simulations involving frictionless grains under isotropic compression.
This suggests that our results are generic, and do not depend,
qualitatively, on the preparation history or on the existence of
friction generated transverse forces between the grains.

Physically, we find that the transition from Gaussian to exponential
force distribution is driven by the localization of force chains as
the applied stress is decreased.  In granular materials, with
particles of similar size, localization is induced by the disorder of
the packing arrangement.  To quantify the degree of localization, we
consider the participation number $\Pi$:
\begin{equation}
\Pi\equiv\left(M\sum_{i=1}^M q_i^2\right)^{-1} \: \: .
\label{pi}
\end{equation}
Here $M$ is the number of contacts between the spheres, $ \langle
Z\rangle =2 M / N$ is the average coordination number, and $N$ is the
number of spheres.  $q_i\equiv {f_i}/{\sum_{j=1}^{M}{f_j}}$, where
$f_i$ is the magnitude of the total force at every contact. From the
definition (\ref{pi}), $\Pi=1$ indicates a limiting state with a
spatially homogeneous force distribution (${q_i}=1/M$, $\forall i$).
On the other hand, in the limit of complete localization, $\Pi\approx
1/M\to 0$ and $M\rightarrow\infty$.

Figure \ref{distri}c shows our results for $\Pi$ versus
 $\sigma$. Clearly, the system is more localized at low stress than at
 high stress.  Initially, the growth of $\Pi$ is logarithmic,
 indicating a smooth delocalization transition.  This behavior is seen
 up to $\sigma\approx$ 2.1 MPa, after which the participation number
 saturates to a higher value:
\begin{equation}
\begin{array}{rcll}
\Pi(\sigma) & \propto & \log (\sigma) & \qquad[ \mbox{$\sigma<$2.1
MPa}]\\ \Pi(\sigma) & \approx & 0.62 & \qquad[ \mbox{$\sigma>$2.1
MPa}]\\
\end{array}
\label{62}
\end{equation}
This behavior suggests that, near the critical density, the forces are
localized in force chains sparsely distributed in space.  As the
applied stress is increased, the force chains become more dense, and
are thus distributed more homogeneously.

How might we expect the participation number to depend upon other
system parameters when the forces are transmitted principally by force
chains?  In an idealized situation, the system has $N_{FC}$ force
chains, each of which has $N_z$ spheres. Each sphere in a force chain
has two major load bearing contacts, which loads must be approximately
equal.  In the lateral directions, roughly four weak contacts are
required for stability. These contacts carry a fraction $\alpha<1$ of
the major vertical load.  All other contacts have $f_i \approx 0$.
Under these assumptions,

\begin{equation}
\Pi = \frac{2}{ \langle Z \rangle}\frac{(1+2\alpha)^2}{(1+2\alpha^2)}
\frac{ N_{FC} N_z}{ N} \leq \frac{2}{ \langle Z \rangle
}\frac{(1+2\alpha)^2}{(1+2\alpha^2)} \:\:\:.
\label{pifc}
\end{equation}
The last inequality becomes an equality {\it iff} all the balls are in
force chains.  From our simulations at large pressure $\alpha\approx
2/5$, so at $\langle Z \rangle \approx 8$, $\Pi \approx 0.62$, which
implies that the system has been completely homogenized.
Although Eq. (\ref{pifc}) is oversimplified, we believe that the
change in slope in Fig. \ref{distri}c is emblematic of the complete
disappearance of well-separated chains.

\begin{figure}
\centerline{
\hbox{
\epsfxsize=5.cm
\epsfbox{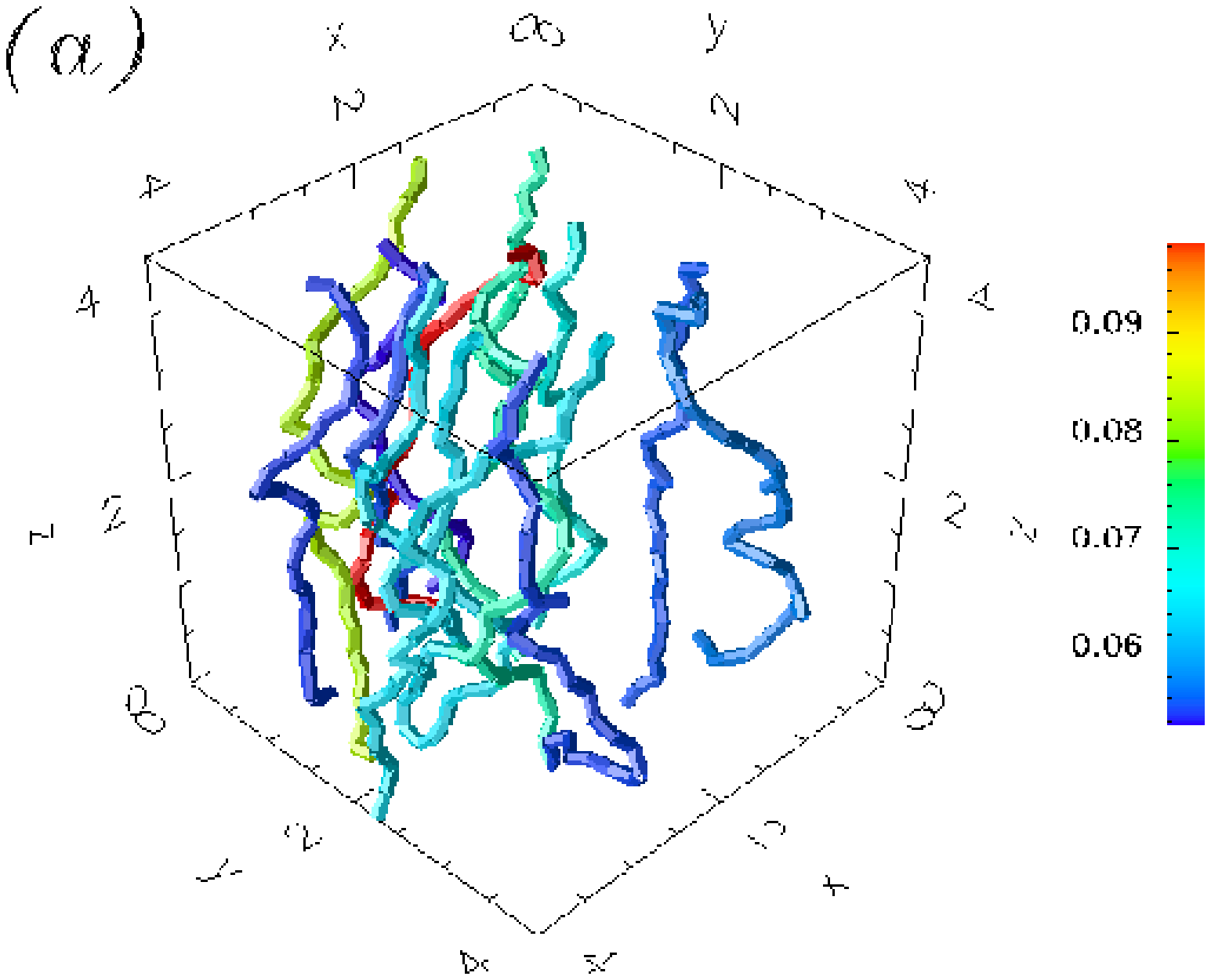}
\epsfxsize=5.cm
\epsfbox{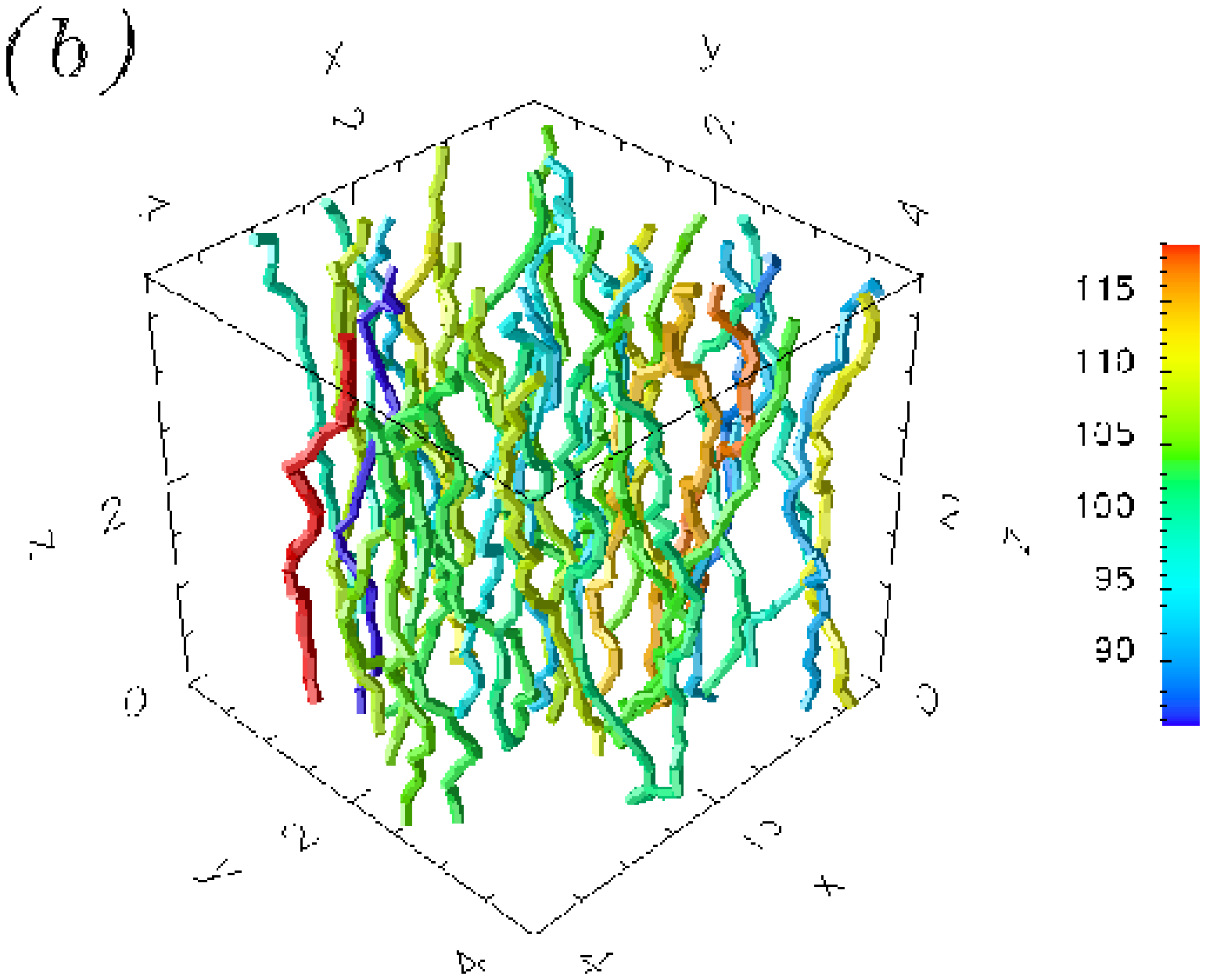}
}
}
\narrowtext
\caption{Example of percolating
force chains for the same system as in Fig. \protect\ref{distri}a: 
(a) near $\phi_c$ and (b) away from $\phi_c$ at large
confining
stress. The color code of the chains is according to the total force
in N carried by the
chains.}
\label{chains}
\end{figure}

The localization transition can be understood by studying the behavior
of the forces during the loading of the sample.  Clearly, visualizing
forces in 3D systems is a complicated task.  In order to exhibit the
rigid structure from the system we visually examine all the forces
larger than the average force; these carry most of the stress of the
system.  The forces smaller than the average are thought to act as an
interstitial subset of the system providing stability to the buckling
of force chains \cite{force_chains,radjai}.  We look for force chains
by starting from a sphere at the top of the system, and following the
path of maximum contact force at every grain.  We look only for the
paths which percolate, i.e., stress paths spanning the sample from the
top to the bottom.  In Fig. \ref{chains} we show the evolution of the
force chains thus obtained for two extreme cases of confining stress.
We clearly see localization at low confining stress: the force-bearing
network is concentrated in a few percolating chains.  At this point
the grains are weakly deformed but still well connected. We expect a
broad force distribution, as found in this and previous studies.  As
we compress further, new contacts are created and the density of force
chains increases. This in turn gives rise to a more homogeneous
spatial distribution of forces, which is consistent with the crossover
to a narrow Gaussian distribution.

{\it Experiments}: Some of the predictions of our numerical study can
be tested using standard carbon paper experiments \cite{chicago},
which have been used successfully in the past to study the force
fluctuations in granular packings.  A Plexiglas cylinder, 5 cm
diameter and varying height (from 3 cm to 5 cm), is filled with
spherical glass beads of diameter $0.8\pm0.05$ mm.  At the bottom of
the container we place a carbon paper with white glossy paper
underneath.  We close the system with two pistons and we allow the top
piston to slide freely in the vertical direction, while the bottom
piston is held fixed to the cylinder.  The system is compressed in the
vertical direction with an Inktron$^{TM}$ press and the beads at the
bottom of the cylinder left marks on the glossy paper. We digitize
this pattern and calculate the ``darkness'' \cite{chicago} of every
mark on the paper.  To calibrate the relationship between the marks
and the force, a known weight is placed on top of a single bead; for
the forces of interest in this study (i.e., from $\approx 0.05$N to 6
N), there is a roughly linear relation between the darkness of the dot
and the force on the bead.

We perform the experiment for different external forces, ranging from
2000 N to 9000 N, and different cylinder heights.  The corresponding
vertical stress, $\sigma$, at the bottom of the cylinder ranges
between 100 KPa and 2.3 MPa (as measured from the darkness of the
dots).
The results of four different measurements are shown in
Fig. \ref{distri}b. For $\sigma$ smaller than $\approx 750$ KPa, the
distribution of forces, $f$, at the bottom piston decays
exponentially:
\begin{equation}
P(f) = \langle f \rangle ^{-1}\exp{[- f/\langle f \rangle ]}, \qquad[
\mbox{ $ \sigma<$ 750 KPa}],
\end{equation}
where $ \langle f \rangle$ is the average force.  When the stress is
increased above 750 KPa there is a gradual crossover to a Gaussian
force distribution as we find in the simulations.  For example, at 2.3
MPa we have
\begin{equation}
P(f) \propto \exp{\left[- k^2 \left(f-f_o\right)^2\right]}, \qquad[
\mbox{$\sigma=$ 2.3 MPa}].
\end{equation}
where $k f_o \approx 1$, and therefore $ \langle f \rangle \approx
f_o$. Similar results have been found in 2D geometries
\cite{behringer}.

{\it Discussion: } In summary, using both numerical simulations and
experiments, we have studied unconsolidated compressible granular
media in a range of pressures spanning almost four decades.  In the
limit of weak compression, the stress vanishes continuously as
$(\phi-\phi_{c})^\beta$, where $\phi_c$ corresponds to RLP or RCP
according to the existence or not of transverse forces between the
grains, respectively. At criticality, the coordination number
approaches a minimal value $Z_c$ (=4 for friction and 6 for
frictionless grains) also as a power law.  Our result $Z_c=6$ agrees
with experimental analysis of Bernal packings for close contacts
between spheres fixed by means of wax \cite{bernal}, and our own
analysis of the Finney packings \cite{finney} using the actual sphere
center coordinates of 8000 steel balls.  However, no similar
experimental study exists for RLP which could be able to confirm
$Z_c=4$.  A critical slowing down--- the time to equilibrate the
system increases near $\phi_c$--- and the emergency of shear rigidity
(to be discussed elsewhere) is also found at criticality.  The
distribution of forces is found to decay exponentially.  The system is
dominated by a fragile network of 
relatively few force chains which span the system.

When the stress is increased away from $\phi_c$ to the point that the
number of contacts has significantly increased from its initial value
$Z_c$ we find: (1) the distribution of forces crosses over to a
Gaussian (2) the participation number increases, and then abruptly
saturates and (3) the density of force chains increases to the point
where it no longer makes sense to describe the system in those terms.
Our simulations indicate that the crossover is associated with a loss
of localization and the ensuing homogenization of the force-bearing
stress paths.

\end{multicols}

\end{document}